\documentclass[twocolumn,
preprintnumbers,prl,nofootinbib,
superscriptaddress
]{revtex4-1}

\usepackage{amsmath,amssymb}
\usepackage[usenames,dvipsnames,svgnames,table]{xcolor}
\usepackage[colorlinks,citecolor=DarkGreen,linkcolor=FireBrick,urlcolor=FireBrick,linktocpage]{hyperref}
\usepackage{graphicx}
\usepackage{times}

\def\Arf{\mathrm{Arf}}
\def\ABK{\mathrm{ABK}}
\def\Cl{\mathrm{Cl}}

\def\Hom{\mathrm{Hom}}
\def\Spin{\mathrm{Spin}}
\def\Pin{\mathrm{Pin}}
\def\DPin{\mathrm{DPin}}
\def\bZ{\mathbb{Z}}
\def\ZZ{\mathbb{Z}}
\def\RP{\mathbb{RP}}
\def\SO{\mathrm{SO}}
\def\OO{\mathrm{O}}

\begin{document}

\title{GSO projections via SPT phases}
\preprint{IPMU-19-0111}

\author{Justin Kaidi}
\author{Julio Parra-Martinez}
\affiliation{Mani L. Bhaumik Institute for Theoretical Physics, \\
Department of Physics and Astronomy,  University of California, Los Angeles, CA 90095, USA}
\author{Yuji Tachikawa}
\affiliation{Kavli Institute for the Physics and Mathematics of the Universe (WPI), \\
 University of Tokyo,  Kashiwa, Chiba 277-8583, Japan}

%\date{\today}

%-----------------------------------------
\begin{abstract}
We point out that the choice of phases in GSO projections can be accounted for by a choice of fermionic SPT phases on the worldsheet of the string.
This point of view not only easily explains why there are essentially two type II theories, but also predicts that there are unoriented type 0 theories labeled by $n$ mod 8,
and that there is an essentially unique choice of the type I worldsheet theory.
We also discuss the relationship between this point of view and the K-theoretic classification of D-branes.
\end{abstract}

\maketitle

\section{Introduction and summary}
The most traditional method of studying superstring theory is via superstring perturbation theory.
In the NS-R formalism, one starts with a worldsheet theory which contains worldsheet spinors which are spacetime vectors.
Then the theory is subjected to a process  known as the Gliozzi-Scherk-Olive (GSO) projection \cite{Gliozzi:1976jf,Gliozzi:1976qd}.
It is well-known that there are various possible choices of consistent GSO projections.
In the standard textbook presentation e.g.~in \cite{Polchinski:1998rr},
allowed GSO projections are determined by imposing various consistency conditions,
such as the modular invariance of torus amplitudes.
That the GSO projections gives consistent results in higher genus amplitudes is not immediately clear in this presentation.

A formulation which works equally well for higher genera was found in \cite{Seiberg:1986by}.
There, it was pointed out that the GSO projection is a summation over the spin structure of the worldsheet, 
and that different GSO projections correspond to different possible phases assigned to spin structures in a way compatible with the cutting and gluing of the worldsheet.
In particular, it was found there that the different signs appearing in type IIA and type IIB GSO projections are given by an invariant of the spin structure  known as the Arf invariant.
The Arf invariant is of order 2, which is closely related to the fact that there are only two type II theories.

Thanks to the developments initiated in condensed matter physics in the last decade, we now have a more physical understanding of this Arf invariant.
Namely, it is the partition function of the low-energy limit of the 1+1d symmetry-protected topological (SPT) phase known as the Kitaev chain \cite{Kitaev:2001kla}.
In general, the low-energy limit of an SPT phase is known as an invertible phase \cite{Freed:2014eja,Freed:2016rqq} and its partition function on a closed manifold is a phase (in the sense of a complex number of absolute value one) which behaves consistently under the cutting and gluing of the spacetime manifold.
Conversely, it is now known that any such consistently-assigned phase is given by the partition function of an invertible theory.
Furthermore, there is now a general classification of possible invertible phases, or equivalently SPT phases, in terms of bordism groups \cite{Kapustin:2014dxa,Freed:2016rqq,Yonekura:2018ufj}.

This means that, with the technology currently available to us, we can now not only understand the consistency of a given GSO projection, but also enumerate all possible GSO projections.
The aim of this letter is to revisit known GSO projections from a modern viewpoint, and possibly find new ones.

For example, the Kitaev chain is known to be compatible with a parity transformation $\Omega$ such that $\Omega^2=(-1)^F$. Also, it is known that eight copies of the Kitaev chain protected by this symmetry are continuously connected to a completely trivial theory \cite{Fidkowski:2009dba}.
In this case, the partition function of the low-energy limit of the Kitaev chain is known as the Arf-Brown-Kervaire (ABK) invariant, and is of order 8.
Let us now consider an unoriented NS-R wordsheet theory with $\Omega^2=(-1)^F$.
Such a worldsheet is said to have a pin$^-$ structure, which is a generalization of the concept of a spin structure to unoriented manifolds (see e.g. Appendix A of \cite{Witten:2015aba} for an introduction).
When we perform the GSO projection, or equivalently when we sum over the pin$^-$ structures,
we can now include $n$ copies of the ABK invariant.
This leads to a series of unoriented type 0 string theories, labeled by $n$ mod 8.
Some of these theories have been discussed in the existing literature \cite{Sagnotti:1995ga,Sagnotti:1996qj,Blumenhagen:1999uy,Blumenhagen:1999bd,Blumenhagen:1999ad,Klebanov:1999pw,Bergman:1999km,Distler:2009ri,Distler:2010an}, but our unified description is new.

We can also ask whether it is possible to modify the GSO projection of the type I theory.
In the type II theory, the left- and right-moving fermions couple to independent spin structures.
This means that the worldsheet fermions of the type I theory have neither pin$^+$ nor pin$^-$ structure;
rather, one needs to consider the spin structure on the orientation double cover of the worldsheet \cite{Distler:2009ri,Distler:2010an}.
We will see below that there are nontrivial invertible phases for this structure, but that they will not lead to any genuinely new type I theory.

We also point out that our viewpoint provides a complementary way to understand the dependence of the K-theoretic classification of D-branes \cite{Witten:1998cd} on the choice of the GSO projection. 
For example, two type II theories differ by the presence of the Arf invariant, or equivalently the Kitaev chain on the worldsheet.
Famously, the Kitaev chain has an unpaired fermionic zero mode on its boundary. 
This explains the fact that the boundary condition for the type IIA non-BPS D9-brane has an unpaired boundary fermion as originally observed in \cite{Horava:1998jy}.
Mathematically, the presence of $n$ boundary fermions corresponds to the existence of the action of the Clifford algebra $\Cl(n)$,
and the K-group $K^{n}(X)$ is defined in terms of unitary bundles with a specified action of $\Cl(n)$ \cite{Karoubi:2008rqk}.
With this observation, we see that the type IIB and type IIA theories have D-branes classified by $K^0(X)$ and $K^1(X)$, respectively.

Similarly, when we have $n$ copies of the ABK invariant, we have $n$ boundary fermions, leading to the existence of the action of the Clifford algebra $\Cl(\pm n)$.
This means that the unoriented pin$^-$ type 0 theory labeled by $n$ mod 8 has D-branes classified by $KO^{+n}(X)\oplus KO^{-n}(X)$.

In the rest of the letter, we will give more details on the points briefly summarized above.
We will work in the lightcone gauge in the NS-R formulation.
A longer version of this letter, filling in many of the details, is forthcoming.

\paragraph{Note added:}
While this work was close to completion,  the authors were informed that E. Witten has an unpublished work with large overlap with this work; two seminars he gave can be found in  \cite{WittenSCGP,WittenStanford}.
The authors thank Kantaro Ohmori and Matthew Heydeman for this information.

\begin{table}[htp]
\begin{center}
\begin{tabular}{|c||c|c|c|c|c|}
\hline
$d$ &   $\mho^d_{\Spin}(pt)$ &    $\mho^d_{\Spin}(B\ZZ_2)$& $\mho^d_{\Pin^-}(pt)$ &  $\mho^d_{\Pin^+}(pt)$ & $\mho^d_{\DPin}(pt)$  
\\\hline\hline
2 &  $\ZZ_2$&$\ZZ_2^{2}$ & $\ZZ_8$&$\ZZ_2$ &$\ZZ_2^2$  
\\\hline
3 & 0& $\ZZ_8$& 0& $\ZZ_2$&$\ZZ_8$ 
 \\\hline
\end{tabular}
\end{center}
\caption{Groups of SPT phases relevant to our analysis. 
The first four columns are classic \cite{KirbyTaylor,Kapustin:2014dxa}.
The last column is new.}
\label{table1}
\end{table}%

\section{Type II strings}
Let us start with type II string theory.
We have eight left-moving fermions $\psi_L^i$ and eight right-moving fermions $\psi_R^i$,
$i=1,\ldots,8$.
We allow for independent spin structures for the left- and right-movers.
As the difference between two spin structures is a $\bZ_2$ gauge field,
we can equivalently say that we have a spin structure and a $\bZ_2$ gauge field.
The chiral GSO projection for the type II theories corresponds to the sum over the left- and right-moving spin structures on the worldsheet.

In general, the global anomaly of a $D$-dimensional fermionic system with symmetry $G$ is controlled by $\mho^{D+1}_\text{Spin}(BG):=\Hom(\Omega_{D+1}^\text{Spin}(BG),U(1))$,
and the possible fermionic SPT phases with symmetry $G$ in $D$ dimensions are classified by $\mho^{D}_\text{Spin}(BG)$.
Here, $\Omega_{d}^\text{Spin}(BG)$ is the bordism group of $d$-dimensional spin manifolds with a $G$-bundle, and $\mho^{d}_\text{Spin}(BG)$ is its Pontryagin dual.
The case relevant for us has $G=\bZ_2$ and $D=2$, for which the groups are given in Table~\ref{table1}.
(We note that in the physics literature our $\mho^d_\text{Spin}(BG)$ is often denoted by $\Omega^d_\text{Spin}(BG)$, 
but this symbol signifies something different known as bordism cohomology for mathematicians, and the authors would like to avoid it.)

We see that the anomaly is characterized by $\mho^3_\text{Spin}(B\bZ_2)=\bZ_8$. 
A single Majorana fermion coupled to two spin structures is known to have an anomaly which is a generator of this $\bZ_8$, 
and, as pointed out in \cite{Tachikawa:2018njr}, the chiral GSO projection is non-anomalous thanks to the fact that we have $10-2=8$ Majorana fermions.

The SPT phases we can add to the worldsheet are classified by $\mho^2_\text{Spin}(B\bZ_2)=\bZ_2^2$, and their partition functions are given by \begin{equation}
(-1)^{n_L \Arf(\sigma_L) + n_R \Arf(\sigma_R)}\label{doubleArf}
\end{equation}
where $\sigma_{L(R)}$ is the left(right)-moving spin structure,
$\Arf(\sigma)$ is the mod-2 Arf invariant,
and $n_{L(R)}=0,1$ label the four choices.
As discussed above, they correspond to four distinct GSO projections possible in type II theories.

We then need to explain why we usually only talk about two type II theories.
For this purpose, we recall that the Arf invariant is the low-energy limit of the Kitaev chain \cite{Kapustin:2014dxa}.
In the language of continuum field theory, this corresponds to the definition %\comment{Citation?}
\cite{Witten:2015aba}
\begin{equation}
e^{i \pi \Arf(\sigma)} = {Z_\text{ferm}(m \gg 0, \sigma)}/{Z_\text{ferm}(m \ll 0, \sigma) }
\end{equation}
where $Z_\text{ferm}(m,\sigma)$ is the partition function of a free massive Majorana fermion of mass $m$; both $\psi_{L,R}$ necessarily couple to the same spin structure $\sigma:=\sigma_R=\sigma_L$. The low-energy limit is taken by sending $|m|\to \infty$, and the denominator is the contribution from a Pauli-Villars regulator. 

In fact this formula holds at finite mass \begin{equation}
e^{i \pi \Arf(\sigma)} = {Z_\text{ferm}(+m, \sigma)}/{Z_\text{ferm}(-m, \sigma) }.
\end{equation}
We further recall that the flip of the sign of the mass term, $m\to -m$,
can be performed by $(\psi_L,\psi_R)\to (\psi_L,-\psi_R)$.
Taking the limit $m\to 0$, we find that a Majorana-Weyl fermion $\psi_R$ has an anomaly under $\psi_R\to -\psi_R$, and generates $(-1)^{\Arf(\sigma_R)}$.

This means that the parity transformation along a single spacetime direction, say $(\psi_L^{i=8},\psi_R^{i=8})\to(-\psi_L^{i=8},-\psi_R^{i=8})$,
produces $(-1)^{\Arf(\sigma_L)+\Arf(\sigma_R)}$, 
i.e.~$n_L=n_R=1$ in \eqref{doubleArf}.
Therefore, there are only essentially two distinct type II GSO projections.
The cases $(n_L,n_R)=(0,0), (1,1)$ are called  type IIB
and the cases $(n_L,n_R)=(0,1), (1,0)$ are called type IIA.

This also explains why T-duality exchanges type IIA and type IIB: 
 T-duality along a spacetime direction implements $(\partial X_R,\psi_R)\to (-\partial X_R,-\psi_R)$ while keeping $(\partial X_L,\psi_L)$ fixed. 
This generates $(-1)^{\Arf(\sigma_R)}$, thus exchanging type IIB and type IIA.

We also note that the Kitaev chain has an unpaired Majorana fermion at its boundary. 
This explains the fact that the boundary condition for the type IIA D9-brane has one boundary fermion \cite{Horava:1998jy}.
The fact that the type IIA D-branes are classified by $K^1(X)$ while the type IIB D-branes are classified by $K^0(X)$ can also be explained from this point of view.
We will discuss this in the context of unoriented pin$^-$ type 0 strings below.

\section{Oriented type 0 strings}
Let us next consider oriented type 0 strings, obtained via diagonal GSO projections \cite{Dixon:1986iz,Seiberg:1986by}.
Here, we assign the same spin structure to both left- and right-moving fermions,
and then sum over this spin structure.
The anomaly is controlled by $\mho^3_\text{Spin}(pt)$ and the invertible phases we can add on the worldsheet are classified by $\mho^2_\text{Spin}(pt)$.
Here $pt$ means the absence of any additional symmetry.

A quick inspection of Table~\ref{table1} shows that there is no anomaly to talk about,
and there are two choices of the invertible phase,
or equivalently two choices of GSO projection.
The invertible phases are simply given by $(-1)^{n\Arf(\sigma)}$ where $n=0,1$.
The case $n=0$ is the oriented type 0B string and the case $n=1$ is the oriented type 0A string.

Compared to type II strings, there is a closed string tachyon in the NSNS sector,
and the RR sector is doubled: 
the type 0B string has two $C$, two $C_{\mu\nu}$, and one non-chiral $C_{\mu\nu\rho\sigma}$ while
 the type 0A string has two $C_\mu$ and two $C_{\mu\nu\rho}$.
Correspondingly, the D-brane spectra are also doubled, and are classified by
$K^0(X)\oplus K^0(X)$ and $K^1(X)\oplus K^1(X)$, respectively. 

\section{Unoriented pin$^-$ type 0 strings}
Let us now move on to a discussion of unoriented versions of type 0 strings.
On unoriented (Wick-rotated) $d$-dimensional spacetime, fermions transform under a double cover of $O(d)$.
There are two distinct choices known as $\Pin^\pm(d)$, distinguished by $\Omega^2=(\pm 1)^F$ where $\Omega$ is a lift of the parity transformation along a single direction.
In our context, this distinction manifests itself, for example, in the spin structure along the boundary of a M\"obius strip.
Going around this boundary is homotopically equivalent to going twice around the unorientable cycle around the crosscap, and thus involves $\Omega^2$.
Therefore, it is automatically in the NS-sector with  pin$^-$ structure,
and in the R-sector with pin$^+$ structure.

We first discuss the pin$^-$ case. 
The system is automatically anomaly-free since $\mho^3_{\Pin^-}(pt)=0$.
We then have the choice of invertible phases on the worldsheet, given by $\mho^2_{\Pin^-}(pt)=\bZ_8$. These invertible phases have been studied previously in \cite{Kapustin:2014dxa,Debray:2018wfz,Turzillo:2018ynq}, and in the condensed matter literature in \cite{Fidkowski:2009dba,Cho:2016xjw,Shiozaki:2016zjg}. 
The partition functions are given by \begin{equation}
e^{(2\pi i/8) n \ABK(\sigma)}\label{ABK}
\end{equation} where $n$ is an integer modulo 8,
and we have $\ABK(\sigma)=\pm1$ on $\RP^2$. 
This in particular means that with $n=4$, $e^{(2\pi i/8)n\ABK(\sigma) }$
assigns the phase $-1$ to $\RP^2$,
and thus maps O9$^-$ to O9$^+$.
More mathematically, this means that \begin{equation}
e^{(2\pi i/8)4\ABK(\sigma) } = (-1)^{\int w_1^2} \label{+-}
\end{equation} where $w_1$ is the first Stiefel-Whitney class of the manifold,
measuring the non-orientability. Theories differing by 4 copies of ABK differ by this same phase in the partition function. 

We also note the following: 
on oriented surfaces, $\ABK(\sigma)=4\Arf(\sigma)$ modulo 8,
which means that the cases $n=0,2,4,6$ are orientifolds of type 0B
and the cases $n=1,3,5,7$ are of type 0A.
The fact that type 0A and 0B theories split into four cases each when we consider unoriented theories was also mentioned in Appendix~F of a recent paper \cite{Stanford:2019vob}.

The Klein bottle admits four pin$^-$ structures, and is obtained by gluing together two copies of $\RP^2$.
As such, the ABK invariants are $(\pm1) + (\pm1) = -2,0,0,+2$.
The Klein bottle amplitude is a trace of $\Omega$ on the closed-string Hilbert space,
and the cases where $\ABK(\sigma)=\pm2$ correspond to the trace of $\Omega$ in the RR-sector,
while those with $\ABK(\sigma)=0$ correspond to the trace in the NSNS-sector.
Therefore, including the phase \eqref{ABK} in the GSO projection only modifies the action of $\Omega$ in the RR-sector, and we have
\begin{equation}
  \Omega^{(n)}_\text{RR} = \Omega^{(0)}_\text{RR} e^{(2\pi i/4) n}\,.
\end{equation}

We find that the NSNS sector contains the tachyon, the metric and the dilaton independent of $n$.
In the RR sector one has 
\begin{equation}
\begin{cases}
C_{\mu\nu}, \quad C'_{\mu\nu} & (n=0,4);\\
C_{\mu}, \quad C_{\mu\nu\rho} & (n=1,5);\\
C, \quad C',\quad C_{\mu\nu\rho\sigma} & (n=2,6);\\
C_{\mu}, \quad C_{\mu\nu\rho} & (n=3,7)
\end{cases}
\end{equation}
where $C_{\mu\nu\rho\sigma}$ is a non-chiral 4-form.
Correspondingly, we can find the following non-torsion D-branes using the standard boundary state formalism:
\begin{equation}
\begin{cases}
D1, D1', D5, D5', D9, D9' & (n=0,4);\\
D0, D2, D4, D6, D8 & (n=1,5);\\
D(-1), D(-1)', D3, D3', D7, D7' & (n=2,6);\\
D0, D2, D4, D6, D8 & (n=3,7).
\end{cases}
\end{equation}
These non-torsion D-branes match the K-theoretic classification given by $KO^{+n}(X)\oplus KO^{-n}(X)$.
This K-theoretic classification of D-branes can also be interpreted as follows.

Let us consider a system of $n$ copies of the Kitaev chain.
When put on a segment, it has $n$ fermion zero modes $\psi_L^a$ on the left boundary and another $n$ fermion zero mode $\psi_R^a$ on the right boundary, $a=1,\ldots,n$.
When we assign the T-transformation $\mathsf{T}\psi_L^a=+\psi_L^a$ and $\mathsf{T}\psi_R^a=+\psi_R^a$,
the anticommutators are $\{\psi_L^a,\psi_L^b\}=+2\delta^{ab}$
and $\{\psi_R^a,\psi_R^b\}=-2\delta^{ab}$ \cite{Fidkowski:2009dba,Witten:2015aba}.
The algebra $\{\psi^a,\psi^b\}=\pm\delta^{ab}$ with $a,b=1,\ldots,n$ 
is known as the Clifford algebra $\Cl(\pm n)$.
This means that when $n$ copies of the ABK invariant are introduced,
there are two types of boundaries, one carrying the action of $\Cl(+n)$ and another carrying the action of $\Cl(-n)$.
Now, the group $KO^{n}(X)$ is defined in terms of orthogonal bundles with an additional action of $\Cl(n)$ \cite{Karoubi:2008rqk}.
Therefore, the D-branes in the unoriented pin$^-$ type 0 string specified by $n$ mod 8
are classified by $KO^{+n}(X) \oplus KO^{-n}(X)$.

\section{Unoriented pin$^+$ type 0 strings}
Let us briefly mention the pin$^+$ case, which has been studied previously under the name type 0' strings \cite{Blumenhagen:1999ns,Bergman:1999km,Angelantonj:1999qg}.
Here the anomaly is characterized by $\mho^3_{\Pin^+}(pt)=\bZ_2$.
As we have eight fermions, the GSO projection is anomaly free.
Then the invertible phases on the worldsheet are classified by $\mho^2_{\Pin^+}(pt)=\bZ_2$.
It is known that the invertible phases are given by \begin{equation}
(-1)^{n \Arf(\hat\sigma)},\quad n=0,1
\end{equation} where $\hat\sigma$ is the spin structure of the orientation double cover of the worldsheet.
When orientable, $\Arf(\hat\sigma)=\Arf(\sigma_L)+\Arf(\sigma_R)$.
This can be generated by the spacetime parity transformation along a single direction as we already discussed,
and therefore there is only an essentially unique way to perform the GSO projection.
This GSO projection removes the closed string tachyon and has an interesting Green-Schwarz cancellation, 
but we do not have anything to add to the discussions given in the references cited above.

\section{Type I string}
We now study the possibility of a new GSO projection for the type I string. It will turn out that there is no new possibility.

In order to explain this, we need to study the spin structure of the orientation double cover. 
Our main interest here is the worldsheet, which is two-dimensional,
but to formulate the bordism group it is useful to work in arbitrary dimensions.
Let us then consider (Wick-rotated) $n$-dimensional manifolds.
When the manifold is oriented, the spin structure of the orientation double cover corresponds to considering $\Spin(n)\times \bZ_2$,
which is a double cover of $\SO(n)\times \bZ_2$.
Therefore, when the manifold is unoriented, we need to consider a double cover \begin{equation}
0 \to \bZ_2 \to G\to \OO(n)\times \bZ_2 \to 0.
\end{equation} 
The precise extension is specified by an element of $H^2(B\OO(n)\times B\bZ_2,\bZ_2)$ and is given  by $w_2+w_1^2+w_1a$, where $w_{1,2}$ are the usual Stiefel-Whitney classes of $\OO(n)$ and $a$ is the generator of $H^1(B\bZ_2,\bZ_2)=\bZ_2$.
We can check that this group $G$ contains $\Spin(n)\times \bZ_2$ as it should,
and also contains both $\Pin^+(n)$ and $\Pin^-(n)$.
This is in some sense a double Pin group, and we denote it by $G=\DPin(n)$.

We need to find $\mho^{2,3}_{\DPin}(pt)$.
This can be computed using the Atiyah-Hirzebruch spectral sequence for twisted spin bordism groups, using the information on the differential $d_2$ recently found in \cite{Thorngren:2018bhj}.
The result is shown in Table~\ref{table1};
we will detail the computation in the upcoming full paper.
The GSO projection is anomaly free, since $\mho^{3}_{\DPin}(pt)=\bZ_8$ and we have eight fermions.
The invertible phases on the worldsheet are classified by $\mho^2_{\DPin}(pt)=\bZ_2^2$,
whose generators are simply \begin{equation}
(-1)^{\int w_1^2}, \qquad (-1)^{\Arf(\hat \sigma)}.
\end{equation}
We already noted in \eqref{+-} that $(-1)^{\int w_1^2}$ simply exchanges O9$^\pm$.
We also already saw that $(-1)^{\Arf(\hat\sigma)}$ can be generated by a spacetime parity transformation along a single direction, and therefore does not  lead to an essentially different GSO projection. Hence one obtains only the usual type I and $\tilde{\rm I}$ strings.

\section{Acknowledgments} 
The authors thank the TASI 2019 summer school at University of Colorado, Boulder,
where JK and JPM were students and YT was a lecturer,
for providing an ideal environment for collaboration;
this letter grew out of a lunchtime conversation there.
The authors also thank Arun Debray, Ryohei Kobayashi and Ryan Thorngren for their help in computing the relevant bordism groups, and Max Metlitski for many useful discussions.
YT is in part supported  by WPI Initiative, MEXT, Japan at IPMU, the University of Tokyo,
and in part by JSPS KAKENHI Grant-in-Aid (Wakate-A), No.17H04837 
and JSPS KAKENHI Grant-in-Aid (Kiban-S), No.16H06335. JPM is supported by the US Department of State through a Fulbright scholarship. JK and JPM thank the Mani L. Bhaumik
Institute for generous support. 

\bibliographystyle{ytphys}
\bibliography{ref}

\end{document}